\documentclass[12pt]{article}

\usepackage{rotating}
\usepackage{amsmath}
\usepackage{amsfonts}
\usepackage{graphicx}
\usepackage{epsfig}
\usepackage{color}
\usepackage{geometry}
\usepackage{pdflscape}

\usepackage{epstopdf}
\usepackage{amssymb}
\graphicspath{ {plots/} }
\usepackage{mathrsfs}
\usepackage{multirow}
\usepackage{psfrag}

\newcommand{\C}{\mathbb{C}}

\newcommand{\fz}{\mathfrak{z}}

\newcommand{\cF}{\mathcal{F}}

\newcommand{\cX}{{\mathcal{X}}}
\newcommand{\cY}{{\mathcal{Y}}}

\newcommand{\be}{\begin{equation}}
\newcommand{\ee}{\end{equation}}
\newcommand{\bea}{\begin{eqnarray}}
\newcommand{\eea}{\end{eqnarray}}
\newcommand{\nn}{\nonumber}

\newcommand{\ed}{\end{document}}

\newcommand{\bi}{\begin{itemize}}
\newcommand{\ei}{\end{itemize}}

\newcommand{\bce}{\begin{center}}
\newcommand{\ece}{\end{center}}

\newcommand{\sG}{\mathscr{G}}

\begin{document}

\title{Unidirectional Reflection and Invisibility in Nonlinear Media with an Incoherent Nonlinearity}

\author{Ali~Mostafazadeh\thanks{Corresponding author, Email Address: amostafazadeh@ku.edu.tr}  ~and Neslihan Oflaz\\[6pt]
Departments of Mathematics and Physics, Ko\c{c} University,\\ 34450 Sar{\i}yer,
Istanbul, Turkey}

\date{ }
\maketitle

\begin{abstract}

We give explicit criteria for the reflectionlessness, transparency, and invisibility of a finite-range potential in the presence of an incoherent (intensity-dependent) nonlinearity that is confined to the range of the potential. This allows us to conduct a systematic study of the effects of such a nonlinearity on a locally periodic class of finite-range potentials that display perturbative unidirectional invisibility. We use our general results to examine the effects of a weak Kerr nonlinearity on the behavior of these potentials and show that the presence of nonlinearity destroys the unidirectional invisibility of these potentials. If the strength of the Kerr nonlinearity is so weak that the first-order perturbation theory is reliable, the presence of nonlinearity does not affect the unidirectional reflectionlessness and transmission reciprocity of the potential. We show that the expected violation of the latter is a second order perturbative effect.
\vspace{2mm}

%\noindent PACS numbers: 03.65.Nk, 42.25.Bs\vspace{2mm}


\end{abstract}

An important difference between real and complex scattering potentials is that the reciprocity in reflection is generally broken for a complex potential. An extreme example is a potential $v(x)$ that is reflectionless only from the left or right. If, in addition, $v(x)$ has perfect transmission property, i.e., the transmitted (left- or right-going) waves are not affected by the presence of $v(x)$, it is said to be unidirectionally invisible \cite{lin,unidir,pra-2013}. The aim of the present letter is to explore the consequences of introducing a weak incoherent nonlinearity on the behavior of a unidirectionally invisible potential.\footnote{Here by ``incoherent'' we mean that the nonlinear term in the wave equation does not depend on the phase of its solutions $\psi$, i.e., it is a function of $|\psi|$.}

Consider an isotropic nonmagnetic medium with translational symmetry along the $y$- and $z$-axes. The interaction of this medium with electromagnetic waves is described by its relative permittivity $\hat\varepsilon(x)$. A normally incident $z$-polarized TE wave that propagates in such a medium has an electric field of the form $\vec E(\vec r,t)=E_0e^{-ikct}\psi(x)\hat e_z$, where $\vec r$ is the position vector, $E_0$ is a constant amplitude, $k$ and $c$ are respectively the wavenumber and the speed of light in vacuum, $\hat e_j$ is the unit vector along the $j$-axis with $j\in\{x,y,z\}$, and $\psi(x)$ solves the Helmholtz equation,
$\psi''(x)+k^2\hat\varepsilon(x)\psi(x)=0$. This is equivalent to the Schr\"odinger equation:
	\be
	-\psi''(x)+v(x)\psi(x)=k^2\psi(x),
	\label{sch-eq}
	\ee
for the optical potential:
	\be
	v(x):=k^2[1-\hat\varepsilon(x)].
	\label{opt-pot}
	\ee

For a nonlinear medium with incoherent nonlinearity, where $\hat\varepsilon$ depends on $|\vec E|$, the role of (\ref{sch-eq}) is played by its nonlinear generalization. For a medium forming a slab that is placed between the planes $x=0$ and $x=L$, this has the form
	\be
	-\psi''(x)+v(x)\psi(x)+\gamma\,\chi(x)\cF(|\psi(x)|)\psi(x) = k^2\psi(x),
	\label{NS}
	\ee
where $\gamma$ is a real coupling constant,
	\begin{align}
	&\chi(x):= \begin{cases}
     	1\, & \  x\in[0,L],\\
     	0\, & \ x\notin[0,L],
     	\end{cases}\\
	&v(x)=\fz\, f(x)\chi(x)=\begin{cases}
     	\fz\, f(x)\, & \  x\in[0,L],\\
     	0\, & \ x\notin[0,L],
     	\end{cases}
	\label{pot}
	\end{align}
$\fz$ is a real coupling constant that we have introduced for future use, $f(x):=\fz^{-1}k^2[1-\hat\varepsilon(x)]$, and $\cF:[0,\infty)\to\C$ is a function representing the nonlinear behavior of the medium.\footnote{We also assume that $\fz$ and $\gamma$ are so that $|f(x)|$ and $|\cF(|\psi|)|$ are bounded by numbers of order 1.} For a Kerr medium, $\cF(|\psi|):=|\psi|^2$.

Equation~(\ref{NS}) is a nonlinear Schr\"odinger equation with a confined nonlinearity. The scattering theory defined by this equation has been of interest in laser physics \cite{pra-2013c,liu-2014} where the behavior of the lasing modes are linked with the corresponding nonlinear spectral singularities (poles of the transmission and reflection coefficients.) The study of the scattering features of nonlinear media modeled by a nonlinear Schr\"odinger equation has a long history \cite{NL-scattering}. The research activity in this subject is however mostly focused on situations where the nonlinearity has an infinite range, see for example \cite{rapedius-2009} where the authors consider the transmission resonances in an extended Kerr medium containing a finite number of delta-function barrier potentials. For a recent review of the physical aspects of nonlinear Schr\"odinger equations involving a complex potential, see \cite{konotop-2016}.

The principal example of a unidirectionally invisible potential is
	\be
	v(x)=\fz\, \chi(x) e^{iKx},
	\label{exp-pot}
	\ee
where $K=\pi/L$,  \cite{lin,unidir-1a,unidir-1b, unidir-1c,unidir-longhi,uzdin,jones-jpa-2012}. This potential turns out to be unidirectionally invisible from the left for $k=2K=2\pi/L$ provided that $|\fz|\ll K^2$. The latter condition is an indication that the unidirectional invisibility of this potential is a first-order pertubative effect. We therefore call it ``perturbative unidirectional invisibility''; it persists provided that the first Born approximation is applicable \cite{pra-2014a}. For sufficiently large values of $|\fz|/k^2$, the potential (\ref{exp-pot}) loses this property \cite{unidir-longhi,pra-2014a}. This turns out to be a common feature of an infinite class of locally periodic potentials of the form
	\be
	v(x)=\fz\, \chi(x)\sum_{n=-\infty}^\infty c_n e^{inK x}.
	\label{gen-pot}
	\ee
Specifically, if $c_{-s}=c_0=0\neq c_s$ for some positive (respectively negative) integer $s$, then the potential (\ref{gen-pot}) displays perturbative unidirectional left (respectively right) invisibility for $k=2 s K=2\pi s/L$, \cite{pra-2014a}. If $c_{\pm s}=0$, the potential is bidirectionally reflectionless, and if in addition $c_0=0$ it is bidirectionally invisible for this wavenumber.

Remarkably the potential (\ref{exp-pot}) displays exact (nonperturbative) unidirectional invisibility for certain values of $\fz/k^2$ that are not necessarily small \cite{jpa-2016}. Other examples of potentials possessing exact unidirectional invisibility are given in \cite{pra-2013,ap-2014,pra-2014b,horsley,longhi-2015,mustafa-pra}.

In this letter we explore the phenomenon of unidirectional invisibility for the scattering processes defined by the nonlinear Schr\"odinger equation (\ref{NS}) with $|\gamma|\ll k^2$. Our starting point is the approach to nonlinear scattering theory that is developed in Ref.~\cite{prl-2013}. This is based on the observation that the scattering solutions of (\ref{NS}) that correspond to right- and left-incident waves are respectively given by
    \bea
    \psi_{k-}(x)&=&
    \left\{\begin{array}{ccc}
    N_-e^{-ikx}&{\rm for}& x<0,\\[3pt]
    \xi_{k}(x)&{\rm for}& 0\leq x\leq L,\\[3pt]
    \displaystyle
    \frac{e^{ik(x-L)}F_+(k)-e^{-ik(x-L)}F_-(k)}{2ik}
    &{\rm for}& x> L,
    \end{array}\right.
    \label{rightpsi}\\[6pt]
	\psi_{k+}(x)&=&
    \left\{\begin{array}{ccc}
    \displaystyle
    \frac{e^{ikx}G_+(k)-e^{-ikx}G_-(k)}{2ik}
    &{\rm for}& x<0,\\[3pt]
    \zeta_{k}(x)&{\rm for}& 0 \leq x \leq L,\\[3pt]
    N_+ e^{ik(x-L)} &{\rm for}& x> L,
    \end{array}\right.
    \label{leftpsi}
    \eea
where  $\xi_k$ and $\zeta_k$ are the solutions of \eqref{NS} in $[0,L]$ satisfying
	\begin{align}
	&\xi_{k}(0)=N_-, &&\xi_{k}'(0)=-ikN_-,
      	\label{ini-1}\\
	&\zeta_k(L)= N_+, &&\zeta'_k(L)=ik N_+,
	\label{ini-2}
	\end{align}
$F_{\pm}$ and $G_{\pm}$ are the Jost functions determined by
	\begin{align}
	&F_{\pm}(k):=\xi'_k(L)\pm i k\xi_k(L),
	&&G_{\pm}(k):=\zeta'_k(0)\pm i k\zeta_k(0),
	\label{Jost}
	\end{align}
and $N_\pm$ are nonzero constants.

In view of \eqref{rightpsi} and \eqref{leftpsi}, we can identify the right/left reflection and transmission coefficients, $R^{r/l}$ and $T^{r/l}$, as follows \cite{prl-2013}:
    \be
    \begin{aligned}
    &R^r=-\frac{e^{-2ikL}F_+(k)}{F_-(k)},~~~~ &&T^r=-\frac{2ik e^{-ikL}N_-}{F_-(k)},\\
	&R^l=-\frac{G_-(k)}{G_+(k)}, &&T^l=\frac{2ik e^{-ikL} N_+}{G_+(k)}.  	
  	\end{aligned}
    \label{R-T}
    \ee
In order to simplify these relations, first we note that $\xi_{k}$ and $\zeta_{k}$ fulfil
	\bea
    \xi_{k} (x)&=&N_{-}e^{-ikx} +\int_{0}^x {\sG} (x,x') [\gamma\cF(|\xi_{k} (x')|) + \fz f(x')]\xi_{k} (x')dx',
    \label{xi}\\
	\zeta_{k} (x)&=&N_{+}e^{ik(x-L)} +\int_{L}^x {\sG} (x,x') [\gamma\cF(|\zeta_{k} (x')|) + \fz f(x')]
	\zeta_{k} (x')dx',
	\label{zeta}
    	\eea
where $\sG(x,x'):=\sin[k(x-x')]/k$ is the Green function for the equation $\psi''+k^2\psi=0$. Next, we
introduce
	\begin{align}
	&\hat\xi_{k}(x):=N_-^{-1} e^{ikx}\xi_{k}(x) ,~~~~~~
    \hat\zeta_{k}(x):=N_+^{-1} e^{ik(L-x)}\zeta_{k}(x) ,
	\label{hat-xi-zeta}\\
	&\cX(x):=\chi(x)\left[\gamma\cF(|N_-\hat\xi_{k}(x)|)+\fz f(x)\right]\hat\xi_{k}(x),
    \label{cX}\\
	&\cY(x):=\chi(x)\left[\gamma\cF(|N_+\hat\zeta_{k}(x)|)+\fz f(x)\right]\hat\zeta_{k}(x),
	\label{cY}
	\end{align}
and use (\ref{Jost}) and (\ref{xi}) -- (\ref{cY}) to express $F_\pm(k)$ and $G_\pm(k)$ in terms of $\cX(x)$ and $\cY(x)$.  Substituting the result in (\ref{R-T}), we find
	\be
    \begin{aligned}
    &R^r=\frac{\tilde{\cX}(2k)}{2ik-\tilde{\cX}(0)},~~~~ &&T^r=\frac{2ik}{2ik-\tilde{\cX}(0)},\\
	&R^l=\frac{\tilde{\cY}(-2k)}{2ik-\tilde{\cY}(0)}, &&T^l=\frac{2ik}{2ik-\tilde{\cY}(0)},
  	\end{aligned}
    \label{R-T-2}
    \ee
where a tilde denotes the Fourier transform; e.g., $\tilde\cX(k):=\int_{-\infty}^\infty e^{-ikx}\cX(x)dx$.

The system is said to be left/right reflectionless (respectively transparent) if $R^{l/r}=0$ (respectively $T^{l/r}=1$). It is said to be left/right invisible if it is both left/right reflectionless and transparent, i.e.,  $R^{l/r}=0$ and $T^{l/r}=1$.  The unidirectional invisibility from left/right refers to situations where the system is left/right invisible but fails to be right/left invisible. In light of (\ref{R-T-2}), we can express the conditions for reflectionlessness, transparency, and invisibility in terms of $\tilde{\cX}$ and $\tilde{\cY}$, as shown in Table~\ref{table1}.
    \begin{table}[!htbp]
    \begin{center}
    \begin{tabular}{|c|c|c|}
    \hline
      & From Right & From Left
    \\
    \hline
    Reflectionlessness & $\tilde{\cX}(2k)=0$ & $\tilde{\cY}(-2k)=0$\\
    \hline
    Transparency & $\tilde{\cX}(0)=0$ & $\tilde{\cY}(0)=0$\\
    \hline
    Invisibility & $\tilde{\cX}(2k)=\tilde{\cX}(0)=0$ & $\tilde{\cY}(-2k)=\tilde{\cY}(0)=0$\\
    \hline
    \end{tabular}
    \vspace{6pt}
    \caption{Conditions for reflectionlessness, transparency, and invisibility for an incident plane wave with wavenumber $k$.}
    \label{table1}
    \end{center}
    \end{table}

A linear medium might violate reciprocity in reflection, but it respects reciprocity in transmission \cite{ahmed-2001,prl-2009}, i.e., $T^l=T^r$. An important feature of nonlinear media is that they can violate the reciprocity in transmission. This is of particular interest in attempts to devise an optical isolator (diode) \cite{diode}.

According to Table~\ref{table1}, we can acquire a more explicit quantitative characterization of unidirectional invisibility for the media described by the nonlinear Schr\"odinger equation (\ref{NS}) if we can calculate $\cX(x)$ and $\cY(x)$. This in turn requires the determination of $\hat\zeta_k$ and $\hat\xi_k$, which in light of (\ref{xi}) -- (\ref{hat-xi-zeta}), satisfy
    \begin{align}
	&\hat\xi_{k}(x)=1+\frac{1}{k}\int_{0}^x \sin[k(x-x')] e^{ik(x-x')}
    [\gamma\cF(|N_-\hat\xi_{k} (x')|) + \fz f(x')]\hat\xi_{k}(x')dx',
	\label{hat-xi-2}\\
	&\hat\zeta_{k}(x)=1+\frac{1}{k}\int_{L}^x \sin[k(x-x')] e^{-ik(x-x')}
    [\gamma\cF(|N_+\hat\zeta_{k}(x')|) + \fz f(x')]\hat\zeta_{k} (x')dx'.
	\label{hat-zeta-2}
    \end{align}
The exact solution of these equations are clearly out of reach. But we can obtain perturbative series expansion for their solution which are reliable for sufficiently weak potentials and nonlinearity profiles, i.e., small $|\gamma|/k^2$ and $|\fz|/k^2$. In what follows we confine our attention to the study of the effects of a weak Kerr nonlinearity on perturbatively invisible potentials of the form (\ref{gen-pot}). In particular, we suppose that $|\fz|/k^2$ and $|\gamma|/k^2$ are so small that we can neglect the quadratic and higher order terms in powers of $|\fz|/k^2$ and $|\gamma|/k^2$. We will then examine the consequences of taking into account the terms of order $\gamma\fz/k^4$ and $\fz^2/k^4$.

To find the first-order perturbative solution of (\ref{hat-xi-2}) and (\ref{hat-zeta-2}), we insert $1$ for the $\hat\zeta_k(x')$ and $\hat\xi_k(x')$ that appear on the right-hand side of these equations. This yields
    \begin{align}
	&\hat\xi_{k}(x)\approx \hat\xi_{k}^{(1)}(x):=1+\frac{1}{k}\int_{0}^x \sin[k(x-x')] e^{ik(x-x')}
    [\gamma\cF(|N_-|) + \fz f(x')]dx',
	\label{hat-xi-app}\\
	&\hat\zeta_{k}(x)\approx \hat\zeta_{k}^{(1)}(x):= 1+\frac{1}{k}\int_{L}^x \sin[k(x-x')] e^{-ik(x-x')}
    [\gamma\cF(|N_+|) + \fz f(x')]dx'.
	\label{hat-zeta-app}
    \end{align}
For a weak Kerr nonlinearity, where $\cF(|\psi|)=|\psi|^2$ and $|\gamma|/k^2\ll1$, Eqs.~\eqref{cX}, \eqref{cY}, \eqref{hat-xi-app}, and \eqref{hat-zeta-app} give
    \begin{align}
    &\cX(x)\approx \cX^{(1)}(x):= \chi(x)\left[\gamma|N_-|^2+\fz f(x)\right],
    \label{cX-app-m}\\
    &\cY(x)\approx \cY^{(1)}(x):=\chi(x)\left[\gamma |N_+|^2 +\fz f(x)\right].
    \label{cY-app-m}
    \end{align}
    \noindent
Using these in \eqref{R-T-2} and neglecting quadratic and higher order terms in $\fz$ and $\gamma$, we find
%    \be
%    \begin{aligned}
%    &R^r\approx %R^{r(1)}:=
%    \frac{\tilde{\cX}^{(1)}(2k)}{2ik},~~~~
%    &&T^r\approx %T^{r(1)}:=
%    1+\frac{\tilde{\cX}^{(1)}(0)}{2ik},\\
%    &R^l\approx %R^{l(1)}:=
%    \frac{\tilde{\cY}^{(1)}(-2k)}{2ik},
%    &&T^l\approx %T^{l(1)}:=
%    1+\frac{\tilde{\cY}^{(1)}(0)}{2ik}.
%    \end{aligned}
%    \label{R-T-2-app}
%    \ee
    \begin{align}
    &R^r\approx \frac{\tilde{\cX}^{(1)}(2k)}{2ik}=
    \gamma|N_-|^2\left( \frac{e^{-2ik}-1}{4k^2}\right)-
    \frac{i\fz}{2k}\int_0^L e^{-2ikx}f(x)dx,
    \label{RR-app-1}\\
    &R^l\approx \frac{\tilde{\cY}^{(1)}(-2k)}{2ik}=
    -\gamma|N_+|^2\left( \frac{e^{2ik}-1}{4k^2}\right)-
    \frac{i\fz}{2k}\int_0^L e^{2ikx}f(x)dx,
    \label{RL-app-1}\\
    &T^r\approx
    1+\frac{\tilde{\cX}^{(1)}(0)}{2ik}=
    1-\frac{i\gamma|N_-|^2L}{2k}-\frac{i\fz}{2k}\int_0^L f(x)dx,
    \label{TR-app-1}\\
    &T^l\approx
    1+\frac{\tilde{\cY}^{(1)}(0)}{2ik}=
    1-\frac{i\gamma|N_+|^2L}{2k}-\frac{i\fz}{2k}\int_0^L f(x)dx.
    \label{TL-app-1}
    \end{align}
These equations describe the scattering features of a slab made out of weak Kerr material in the first Born approximation. The following are simple consequences of the last two of these equations.
    \begin{enumerate}
    \item For the left- and right-incident waves of identical amplitude, i.e., $N_+=N_-$, $T^l\approx T^r$. Therefore, the expected violation of transmission reciprocity does not manifest itself in the first-order perturbative expression for the transmission amplitudes.
    \item Because $\gamma$ is a real parameter, the presence of the nonlinearity can only change the phase of $T^{l/r}$. In particular a weak Kerr nonlinearity does not affect the left and right transmission coefficients $|T^{l/r}|^2$.
    \end{enumerate}

Next, we employ (\ref{RR-app-1}) -- (\ref{TL-app-1}) to compute the reflection and transmission amplitudes of a potential of the form \eqref{exp-pot}. This gives
%\begin{align}
%&\tilde{\cX}^{(1)}(k)= i \gamma|N_-|^2 \left(\frac{e^{-i k L}-1}{k} \right)+i\fz \left(\frac{e^{-i (k-K) L}-1}{k'-K} \right),
%\label{cX-app}\\
%&\tilde{\cY}^{(1)}(k)= i \gamma|N_+|^2  \left(\frac{e^{-i k L}-1}{k} \right)+i\fz \left(\frac{e^{-i (k-K) L}-1}{k-K} \right).
%\label{cY-app}
%\end{align}
%Substituting these equations in (\ref{R-T-2-app}), we obtain
    \begin{align}
    R^r&\approx \gamma |N_-|^2 \left(\frac{e^{-2ikL}-1}{4 k^2}\right)
    +\fz \left[\frac{e^{-i(2k-K)L}-1}{2 k(2k-K)}\right] ,
    \label{Rr-exp}\\[4pt]
    R^l&\approx -\gamma  |N_+|^2 \left(\frac{ e^{ 2 i k L}-1}{4 k^2}\right)
    -\fz \left[\frac{e^{i (2k+K)L}-1}{2k(2k+K)}\right] ,
    \label{Rl-exp}\\[6pt]	
    T^r&\approx 1-\frac{i\gamma|N_-|^2L}{2k}
     - \fz \left(\frac{e^{iKL} -1}{2kK}\right) ,
    \label{Tr-exp}\\[4pt]
    T^l &\approx 1-\frac{i\gamma|N_+|^2L}{2k}-\fz \left(\frac{e^{i K L}-1}{2 k K}\right) .
    \label{Tl-exp}
    \end{align}

Equations \eqref{Rr-exp} and \eqref{Rl-exp} imply that the system we consider is
unidirectionally reflectionless from the right (respectively left) for a wavenumber $k$ provided that there is a positive integer $m$ such that $k=-K/2=\pi m/L$ (respectively $k=K/2=\pi m/L$). These conditions preserve their form in the absence of nonlinearity. Therefore, to the first order of perturbation theory, nonlinearity does not affect the unidirectional reflectionlessness of the system. Notice also that whenever these conditions hold, the third term on the right-hand side of (\ref{Tr-exp}) and (\ref{Tl-exp}) vanishes. For a linear medium this implies $T^{l/r}\approx 1$. Hence a linear medium satisfying one of these conditions is bidirectionally transparent and invisible from right or left. This is not the case when we take into account the contribution of nonlinearity, because the second term on right-hand side of (\ref{Tr-exp}) and (\ref{Tl-exp}) makes $T^{l/r}$ differ from unity.

We can easily generalize Eqs.~\eqref{Rr-exp} -- \eqref{Tl-exp} to the locally periodic potentials \eqref{gen-pot}, which are sums of potentials of form \eqref{exp-pot} with $K \rightarrow K n$ and $\fz \rightarrow \fz c_n$. The result is
    \begin{align}
    R^r& \approx \gamma |N_-|^2\left(\frac{e^{-2 i k L}-1}{4 k^2}\right)+
    \fz  \sum _{n=-\infty }^{\infty } c_n\left[\frac{e^{-i (2k- nK) L}-1}{2 k (2k -nK)}\right],
    \label{Rr-gen}\\
    R^l&\approx -\gamma  \left|N_+\right|^2 \left(\frac{ e^{2 i k L}-1}{4 k^2}\right)
    -\fz \sum _{n=-\infty }^{\infty }  c_n\left[\frac{e^{i (2k+ nK) L}-1}{2 k (2k + nK)}\right],
    \label{Rl-gen}\\
	T^r&\approx 1-\frac{i\gamma|N_-|^2L}{2k} -\fz \sum _{n=-\infty }^{\infty }
    c_n\left(\frac{e^{i  n K L}-1}{2 n k K}\right),
    \label{Tr-gen}\\
    T^l& \approx 1-\frac{i\gamma|N_+|^2L}{2k} -\fz \sum _{n=-\infty }^{\infty }
    c_n\left(\frac{e^{i  n K L}-1}{2 n k K}\right),
    \label{Tl-gen}
    \end{align}
where $K>0$.

Let $\ell$ denote the period of the potential (\ref{gen-pot}), i.e., $\ell:=2\pi/K$, and suppose that there are positive integers $m$ and $s$ such that $L= m\ell$ and
    \be
    k=\frac{sK}{2}=\frac{\pi s}{\ell}=\frac{m s\pi}{L}.
    \label{k=}
    \ee
Then Eqs.~\eqref{Rr-gen} -- \eqref{Tl-gen} reduce to
    \begin{align}
    R^r& \approx -\frac{i \fz Lc_s}{2k},\quad
    &&R^l\approx  -\frac{i \fz Lc_{-s}}{2k},
    \label{R2simp}\\
     T^r&\approx 1-\frac{i\gamma|N_-|^2L}{2k} - \frac{i \fz Lc_0}{2k}, \quad
     &&T^l \approx  1-\frac{i\gamma|N_+|^2L}{2k}-\frac{i \fz Lc_0}{2k}.
    \label{T2simp}
    \end{align}
If $c_s=0\neq c_{-s}$ (respectively $c_s\neq0=c_{-s}$), the potential \eqref{gen-pot} is unidirectionally reflectionless from right (respectively left) for the wavenumber (\ref{k=}). Again the presence of nonlinearity does not affect the reflectionlessness of the potential (in the first order of perturbative theory), but it obstructs its transparency and invisibility even when $c_0=0$. Notice, however, that whenever $c_0$ takes a real value, in particular when $c_0=0$, the contribution of nonlinearity to the transmission amplitudes amounts to a change of their phase. Therefore, for the wavenumber (\ref{k=}), the potential has unit transmission coefficient; $|T^{l/r}|^2\approx 1$.

Next, we examine the effect of including the terms of order $\gamma\fz$ and $\fz^2$. This requires the calculation of \eqref{cX} and \eqref{cY} using second order perturbation theory. In view of \eqref{cX} and \eqref{cY}, we can express the result in the form
	\begin{align}
	&\cX(x)\approx \cX^{(2)}(x):= \chi(x)\left[\gamma\cF(|N_-\hat\xi_{k}^{(1)}(x)|)+\fz f(x)\right]\hat\xi_{k}^{(1)}(x),
	\label{cX-app-2}\\
	&\cY(x)\approx \cY^{(2)}(x):=\chi(x)\left[\gamma\cF(|N_+\hat\zeta_{k}^{(1)}(x)|)+\fz f(x)\right]\hat\zeta_{k}^{(1)}(x).
	\label{cY-app-2}
	\end{align}
Using these relations in (\ref{R-T-2}), we have
    \be
    \begin{aligned}
    &R^r\approx R^{r(2)}:=\frac{\tilde{\cX}^{(2)}(2k)}{2ik} - \frac{\tilde{\cX}^{(1)}(2k) \tilde{\cX}^{(1)}(0) }{4 k^2},
    &&T^r\approx T^{r(2)}:= 1+\frac{\tilde{\cX}^{(2)}(0)}{2ik} - \frac{[\tilde{\cX}^{(1)}(0)]^2 }{4 k^2},\\
    &R^l\approx R^{l(2)}:=\frac{\tilde{\cY}^{(2)}(-2k)}{2ik} - \frac{\tilde{\cY}^{(1)}(-2k) \tilde{\cY}^{(1)}(0)}{4 k^2},
    &&T^l\approx T^{l(2)}:=1+\frac{\tilde{\cY}^{(2)}(0)}{2ik}  - \frac{[\tilde{\cY}^{(1)}(0)]^2}{4 k^2}.
    \end{aligned}
    \label{R-T-2-app-2}
    \ee
The explicit form of these relations for the potentials of the form \eqref{gen-pot} is quite complicated. We therefore confine our attention to the case of a slab of thickness $L= m\ell$ and incident waves with wavenumber (\ref{k=}), so that for $c_s=0\neq c_{-s}$ (respectively $c_s\neq0= c_{-s}$) the slab is right- (respectively left-) reflectionless up to the first order in $\fz$ and $\gamma$. Under these conditions (\ref{R-T-2-app-2}) gives
    \be
    \begin{aligned}
    R^{(r/l)} &\approx \hat\fz\, R^{(r/l)}_\fz+ \hat\gamma \hat\fz\, R^{(r/l)}_{\gamma\fz}+
    \hat\fz^2\,  R^{(r/l)}_{\fz^2},\\
    T^{(r/l)} &\approx 1 +\hat\fz\, T^{(r/l)}_\fz+ \hat\gamma \,T^{(r/l)}_\gamma+
    \hat\fz\hat\gamma\, T^{(r/l)}_{\gamma\fz}+ \hat\fz^2  \,T^{(r/l)}_{\fz^2},
    \end{aligned}
    \label{2nd-order}
    \ee
where
    \begin{align}
    &\hat\fz:=\frac{\fz}{k^2},~~~~~~~~~\hat\gamma:=\frac{\gamma}{k^2},
    &&R^r_{\fz}=-\frac{ i \pi m s c_s}{2} , && R^l_\fz=-\frac{i\pi m s c_{-s}}{2},
    \label{2ndfirst1}\\
    & T^{l/r}_{\fz}=-\frac{i\pi m s c_0}{2},
    &&T^r_{\gamma}=-\frac{i\pi m s |N_-|^2}{2}, && T^l_\gamma=-\frac{i\pi m s |N_+|^2}{2},
   \label{2ndfirst2} \end{align}
    \begin{align}
    \label{Rr2}
    R^r_{\gamma \fz}=& \frac{i\pi m s|N_-|^2}{16}
    \Big[ 3 c_{-s}+ c_{-s}^* + 2(6c_0-c_0^*) + 6(-1+i\pi m s) c_s+ c_s^*  -2s\sum_{n\neq0,\pm s}\frac{3c_n}{n-s}\Big] ,\\
    R^l_{\gamma \fz}=& \frac{i\pi m s|N_+|^2}{16}
    \Big[ 6(-1+i\pi ms ) c_{-s}+ c_{-s}^* + 2(6c_0-c_0^*)+ 3 c_s+ c_s^*  +2s\sum_{n\neq0,\pm s} \frac{3c_n}{n+s} \Big],\\
    \nn
    R^r_{\fz^2}=&\frac{i\pi m s}{16}
    \Big[ 2 c_{-s}c_0 +4 c_{-s}c_s + c_{-s}c_{2s}  + 4 c_0^2 +4(-1+ i\pi ms)c_0c_s +
    4 c_s^2 \\
    & +2s\sum_{n\neq0,\pm s}\Big(-\frac{2 c_0 c_n}{n-s} +\frac{2 c_s c_n}{n}+ \frac{s c_{-n} c_{n+s}}{n(n+s)}\Big)\Big],\\
    \nn R^l_{\fz^2}=& \frac{i\pi m s }{16}
    \Big[ c_{-2s}c_s+4 c_{-s}^2 +4(-1+ i \pi m s) c_{-s}c_{0} -4 c_{-s}c_s  + 4 c_0^2  +
    2 c_0 c_s \\
    & +2s\sum_{n\neq0,\pm s}\Big( -\frac{2 c_{-s} c_n}{n}+\frac{2 c_0 c_n}{n+s}+ \frac{s c_{-n} c_{n-s}}{n(n-s)}\Big)\Big],
    \end{align}
    \begin{align}
    \label{Trgammaz}
    T^r_{\gamma \fz}=&\frac{i\pi m s|N_-|^2}{16}
    \Big[2 c_{-s}^* + 2[(3-i \pi m s)c_0+ (1- i \pi m s)c_0^*]- 2(3 c_s+ 2 c_s^*) -2s\sum_{n\neq0,\pm s} \frac{c_n+c_n^*}{n}  \Big],\\
    \label{Tlgammaz}
    T^l_{\gamma \fz}=&\frac{i\pi m s|N_+|^2}{16}
    \Big[-2(3 c_{-s}+2 c_{-s}^*)+ 2[(3-i \pi m s)c_0+(1-i \pi m s)c_0^*]+ 2 c_s +
    2s\sum_{n\neq0,\pm s} \frac{c_n+ c_n^*}{n} \Big],\\
    \nn T^{l/r}_{\fz^2}=&\frac{i\pi m s}{16}
    \Big[-2 c_{-s}c_0 + (1-2 i \pi m s) c_{-s}c_s +2 c_{-s}c_{2s}  + 2 (1-i \pi m s) c_0^2 -2 c_0c_{s} - c_s^2  \\
    & \label{Tlrz2}
     +2s\sum_{n\neq0,\pm s}\Big( \frac{ c_{-s} c_{n+s}}{n}-\frac{ c_s c_n}{n+s}-\frac{s c_{n} c_{-n}}{n(n-s)}\Big)\Big].
    \end{align}
According to \eqref{Trgammaz} and \eqref{Tlgammaz}, $T^{l/r}_{\gamma \fz}$ do not generally coincide even when $N_-=N_+$. This is an indication that the violation of reciprocity in transmission is a  second-order effect in perturbative theory.\footnote{Note also that the contribution of the nonlinearity to both the reflection and transmission amplitudes is linear in the intensity of the incident wave.}

In the remainder of this article, we examine some of the consequences of (\ref{2nd-order}) -- (\ref{Tlrz2}) for a potential of the form  \eqref{pot} with
    \begin{equation}
    \label{ex1}
    f(x)= c_{-6}\; e^{-6 i K x}+c_{-2}\; e^{- 2 i K x }+c_{4}\; e^{4 i K x },
    \end{equation}
where $c_{-6},c_{-2}$, and $c_{4}$ are nonzero complex numbers, and $K$ is an integer multiple of $2\pi/L$. The first-order perturbative calculation of the reflection and transmission amplitudes shows that this potential is unidirectionally reflectionless from the right (respectively left) for $k=K$ and $k=3K$ (respectively $k=2K$), and bidirectionally reflectionless for $k= n K/2$ for positive integers $n$ other than $2,4$, and $6$. This statement is independent of the values of the coupling constants $c_{-2},c_{4}$, and $c_{-6}$. %The opposite is the case when we take into account the second order perturbative corrections.

\begin{figure}[!ht]
	\centering
   	\includegraphics[width=0.5\textwidth]{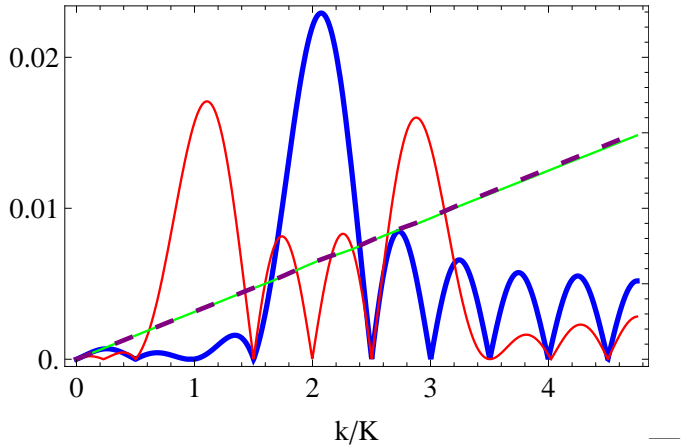}|
	\caption{$|R^r|$ (thick solid blue curve), $|R^l|$ (thin solid red curve), $|T^r-1|$ (thick dashed purple curve), $|T^l-1|$ (thick dashed green curve) as a function of $k/K$ for a potential of the form \eqref{pot} with $f(x)$ given by \eqref{ex1}, (\ref{as=1}), and (\ref{specifics1}). The difference between $|T^{r/l}-1|$ are too small to be seen in the graph.}
	\label{fig1}
	\end{figure}

Figure~\ref{fig1} shows the plots of $|R^{r/l}|$ and $|T^{r/l}-1|$ for a potential of the form (\ref{ex1}) with
	\begin{align}
	&c_{-2}=0.50, && c_{4}=0.35, && c_{-6}=-0.15,
	\label{as=1}\\
	&\hat\gamma |N_\pm|^2= 10^{-3} ,
	&&\hat\fz= 10^{-2} ,
	&& KL= 2\pi~({\rm i.e.,}~ m=1).
	\label{specifics1}
	\end{align}
For larger values of $k$ the transmission amplitudes differ from unity more appreciably, and the violation of transparency and therefore unidirectional invisibility of the potential for $k=K, 2K,$ and $3K$ is stronger.

\begin{figure}[!ht]
	\centering
   	\includegraphics[width=0.5\textwidth]{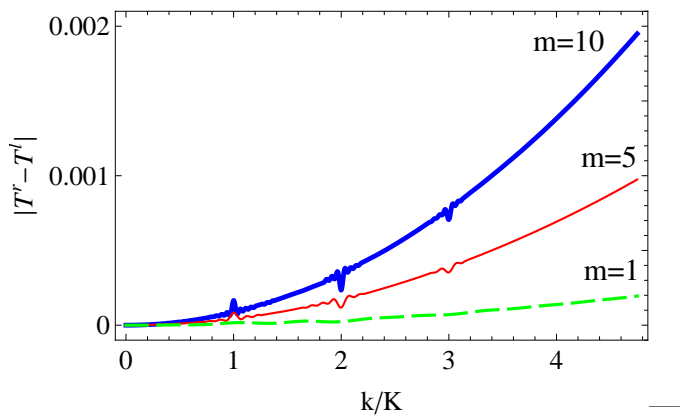}|
	\caption{Plots of $|T^r-T^l|$ as a function of $k/K$ for a potential of the form \eqref{pot} with $f(x)$ given by \eqref{ex1}, (\ref{as=1}), (\ref{specifics1}), and $m=1,5$, and $10$. }
	\label{fig2}
	\end{figure}

Figure~\ref{fig2} shows plots of $|T^r-T^l|$ for the same configuration as in Fig.~\ref{fig1} and $m=1,5,$ and $10$. It reveals the violation of transmission reciprocity that arises due to terms of order $\gamma\fz$. As seen from these plots the violation of transmission reciprocity is more pronounced for larger values of $m$.

Although it is not clear from Fig.~\ref{fig1}, the presence of nonlinearity obstructs the bidirectional invisibility of the potential for $k=nK/2$ and $n\neq 2,4,6$. This is also a second order perturbative effect. For example for $k=4K$, we have $|R^l-R^r|>3\times 10^{-4}$. Because this is much larger than $|N_\pm|^2\hat\gamma\hat\fz=10^{-5}$, the second order perturbation theory gives reliable values for $R^{r/l}$. Using these we find $|R^l/R^r|\approx 4.79$, which is a clear indication of the violation of bidirectional reflectionlessness at $k=4K$.

In summary, we have derived analytic formulas for the reflection and transmission amplitudes of a finite-range potential in the presence of a incoherent nonlinearity that is confined to the range of the potential. These in turn yield simple criteria for the reflectionlessness, transparency, and the invisibility of the nonlinear medium modeled using such a potential. For a weak Kerr nonlinearity, the expression for the reflection and transmission amplitudes simplify considerably. We have used them to show that the nonlinearity destroys the transparency and therefore the unidirectional invisibility of the medium. This stems from the fact that a weak incoherent nonlinearity contributes to the phase of the transmission amplitudes $T^{r/l}$. This contribution is the same for both the left- and right-incident waves, if the nonlinearity is so weak that the first-order perturbation theory is reliable. Therefore the expected nonreciprocal transmission property of the medium does not arise in the first order perturbative calculations. By examining the second-order perturbative expression for $T^{r/l}$ we have shown that nonreciprocal transmission is indeed a second order effect.

\subsection*{Acknowledgments} This work has been supported by  the Scientific and Technological Research Council of Turkey (T\"UB\.{I}TAK) in the framework of the project no: 114F357, and by the Turkish Academy of Sciences (T\"UBA).

\end{document}